\documentclass{aimc2026}

\usepackage[utf8]{inputenc} 
\usepackage[T1]{fontenc}    
\usepackage{url}            
\usepackage{booktabs}       
\usepackage{amsfonts}       
\usepackage{nicefrac}       
\usepackage{microtype}      
\usepackage{graphicx}       
\usepackage{amsmath}
\usepackage{amssymb}

\usepackage[table]{xcolor}
\usepackage{titlesec}
\usepackage{caption}
\usepackage{tcolorbox}
\tcbuselibrary{skins}

\usepackage{hyperref}       
\hypersetup{
  colorlinks = true,
  linkcolor  = blue!60!black,
  citecolor  = blue!55!black,
  urlcolor   = teal!80!black,
  pdftitle   = {nnAudio 2: Overcoming Dynamic Compilation Barriers and Transform Inconsistencies},
  pdfauthor  = {Abhinaba Roy, Junyi Liang, and Dorien Herremans}
}
\usepackage{cleveref}       

\definecolor{secblue}{HTML}{1F3A5F}      
\definecolor{tldrblue}{HTML}{2C5282}     

\tcbset{
  tldr/.style={
    enhanced,
    colback=tldrblue!4,
    colframe=tldrblue!85,
    boxrule=0.6pt,
    arc=5pt,
    left=8pt, right=8pt, top=5pt, bottom=5pt,
    title={\bfseries\small In a Nutshell},
    fonttitle=\bfseries,
    coltitle=white,
    attach boxed title to top left={yshift=-2.2mm, xshift=8pt},
    boxed title style={colback=tldrblue, arc=2pt, boxrule=0pt},
    before skip=6pt, after skip=10pt,
    fontupper=\small
  }
}
\newtcolorbox{nutshell}{tldr}

\titleformat{\section}
  {\normalfont\large\bfseries\color{secblue}}
  {\thesection.}{0.6em}{}
\titleformat{\subsection}
  {\normalfont\normalsize\bfseries\color{secblue}}
  {\thesubsection}{0.6em}{}
\titleformat{\subsubsection}
  {\normalfont\small\bfseries\itshape\color{secblue!85!black}}
  {\thesubsubsection}{0.6em}{}
\titlespacing*{\section}      {0pt}{1.6ex plus .2ex minus .2ex}{0.6ex}
\titlespacing*{\subsection}   {0pt}{1.2ex plus .2ex minus .2ex}{0.3ex}
\titlespacing*{\subsubsection}{0pt}{1.0ex plus .2ex minus .2ex}{0.3ex}

\makeatletter
\global\let\@aimcthanks\@empty
\renewcommand{\@noticestring}{}
\makeatother

\title{nnAudio 2: Overcoming Dynamic Compilation Barriers and Transform Inconsistencies}

\author{%
  Abhinaba Roy \\
  Singapore University of Technology and Design \\
  \texttt{abhinaba\_roy@sutd.edu.sg} \\
  \And
  Junyi Liang \\
  Singapore University of Technology and Design \\
  \texttt{liangjunyi010@gmail.com} \\
  \And
  Dorien Herremans \\
  Singapore University of Technology and Design \\
  \texttt{dorien\_herremans@sutd.edu.sg} \\
}

\begin{document}

\twocolumn[%
  \maketitle
  \begin{nutshell}
    \textbf{nnAudio 2} is a conservative, targeted modernization of the
    nnAudio audio feature-extraction toolbox. We make the STFT and iSTFT
    classes compatible with \texttt{torch.jit.script}; replace the silent
    inverse-STFT failure under non-uniform frequency scales with an explicit
    runtime error, restricting reliable inversion to
    \texttt{freq\_scale=`no'}; restore CFP under modern SciPy; and make VQT
    reduce to CQT at $\gamma = 0$. We also introduce iCQT, a differentiable
    Landweber-based inverse CQT that reconstructs waveforms at $>$30\,dB SNR.
    The updated codebase passes the full repository test suite under a current
    Python and PyTorch environment. Code:
    \url{https://github.com/AMAAI-Lab/nnAudio2}.
  \end{nutshell}
  \begin{abstract}
nnAudio is an open-source audio feature extraction toolbox for deep learning, but its use in current environments is hindered by TorchScript incompatibilities, inverse-transform edge cases, and dependency drift. We present a targeted modernization for modern PyTorch and scientific Python. We resolve TorchScript compilation failures in STFT and iSTFT by removing dynamic state mutation and module construction from scripted code paths and tightening argument handling in inverse-related helpers. We clarify inverse-STFT behavior by restricting reliable inversion to the uniform-bin setting (\texttt{freq\_scale=`no'}) and raising explicit runtime errors for unsupported frequency scales, preventing silently degraded reconstructions. We restore CFP compatibility with modern SciPy and ensure VQT reduces to CQT when $\gamma = 0$. Regression tests cover the new STFT/iSTFT behaviors, and the updated codebase passes the full repository test suite in a modern Python environment. These improvements provide a more robust foundation for differentiable audio analysis in research and deployment.
  \end{abstract}%
]

\aimcnotice 

\section{Introduction}
\label{sec:intro}

Differentiable audio front-ends are a now-standard ingredient in deep learning systems for music. Whether the goal is automatic transcription, source separation, melody extraction, or generative modeling, models routinely consume short-time Fourier transform (STFT), Mel, or constant-Q transform (CQT) features computed on-the-fly from raw waveforms. The original nnAudio toolbox \citep{cheuk2020nnaudio} popularized this approach in PyTorch by expressing each transform as a 1D convolution whose kernels are exposed as ordinary \texttt{nn.Module} parameters. This formulation offers two practical benefits: spectrograms are produced inside the computation graph, removing the need to pre-render and store features on disk; and the Fourier-like bases themselves can, in principle, be made trainable via gradient descent.

nnAudio became a common dependency in music information retrieval pipelines, supporting STFT, Mel spectrograms, MFCCs, CQT (in two algorithmic variants), the more recent variable-Q transform (VQT) \citep{schorkhuber2014matlab}, the gammatone front-end used in auditory modeling, and the combined frequency and periodicity (CFP) representation of \citet{su2015combining}. Over time, however, the maintenance load of the toolbox has outpaced its single-maintainer model.\footnote{The upstream repository carries an explicit ``maintainers wanted'' notice.} Three forms of decay have accumulated:
(i)~the STFT/iSTFT modules use Python idioms that current versions of \texttt{torch.jit.script} reject, blocking deployment via TorchScript;
(ii)~the inverse STFT (iSTFT) silently returns degraded reconstructions in some configurations the API technically permits;
and (iii)~auxiliary modules drift against modernized versions of upstream scientific Python packages, with CFP failing to construct under recent SciPy releases that have reorganized window functions into the \texttt{scipy.signal.windows} subpackage.

These are not headline algorithmic problems, but they are exactly the kind of problems that quietly erode the usefulness of a research toolbox. A user who cannot script their model loses access to a major optimization and deployment path. A user who reconstructs a waveform from an inverted log-frequency spectrogram and obtains a quiet, plausible-sounding but incorrect signal may not notice for some time, and may publish results that are subtly wrong. A user who upgrades SciPy and finds that an entire feature class throws \texttt{AttributeError} at import simply switches libraries.

In this paper we present \emph{nnAudio 2}\footnote{Source code: \url{https://github.com/AMAAI-Lab/nnAudio2}.}, a targeted modernization of the toolbox aimed at exactly these failure modes. Our contributions are deliberately conservative in scope: we do not propose new transforms, new training procedures, or new benchmarks. Instead, we make five concrete changes that preserve nnAudio's API surface where possible and tighten its semantic and engineering contracts where necessary:

\begin{enumerate}
  \item We make the STFT and iSTFT classes compatible with \texttt{torch.jit.script} on current PyTorch, by removing dynamic state mutation and submodule construction from scripted code paths and by giving inverse-related helper arguments concrete, statically inferable types.
  \item We restrict reliable inversion to the uniform-bin setting (\texttt{freq\_scale=`no'}) and raise informative runtime errors for non-uniform frequency scales rather than producing silently degraded reconstructions.
  \item We restore CFP compatibility with modern SciPy by importing \texttt{blackmanharris} from \texttt{scipy.signal.windows} rather than from the top level of \texttt{scipy.signal}, where the function is no longer re-exported.
  \item We ensure that VQT reduces to CQT behavior when its bandwidth-offset parameter $\gamma$ is zero, eliminating a divergence from the established mathematical relationship between the two transforms.
  \item We introduce iCQT, a Landweber-based inverse CQT module that reconstructs waveforms from complex CQT coefficients with $>$30 dB SNR while preserving end-to-end differentiability.
\end{enumerate}

We validate these changes through (a)~regression tests that exercise the new STFT/iSTFT contracts, (b)~end-to-end execution of the existing repository test suite under a current Python environment, and (c)~a TorchScript scripting smoke test that confirms forward and inverse STFT can be compiled and executed.

The remainder of the paper is organized as follows. Section~\ref{sec:related} situates nnAudio relative to other GPU-accelerated audio front-ends. Section~\ref{sec:background} reviews the relevant aspects of TorchScript, the STFT/iSTFT pair as implemented in nnAudio, and the CFP and VQT transforms. Section~\ref{sec:issues} characterizes the four classes of problem we observed in the upstream code. Section~\ref{sec:fixes} describes our fixes. Section~\ref{sec:validation} reports validation results. Section~\ref{sec:discussion} discusses implications and limitations, and Section~\ref{sec:conclusion} concludes.

\begin{figure*}[!t]
  \centering
  \includegraphics[width=0.92\textwidth]{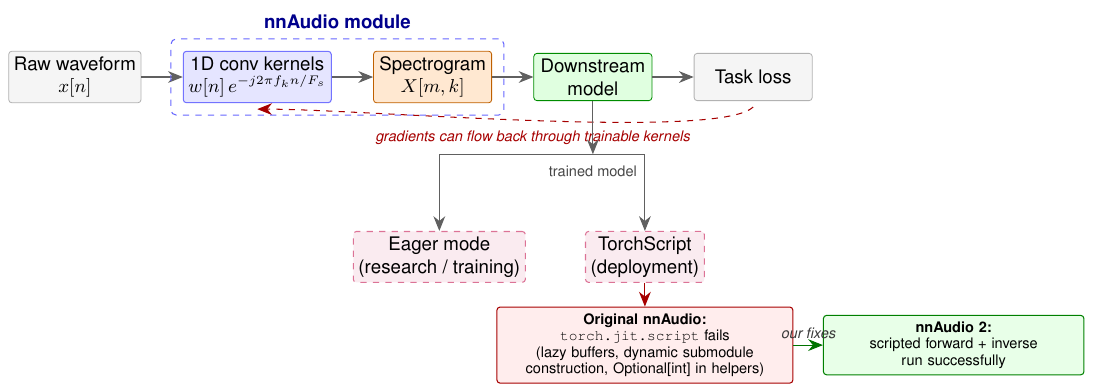}
  \caption{nnAudio expresses each spectrogram as a 1D convolution whose
    kernels are parameters of an \texttt{nn.Module}, allowing gradients
    to flow back into the analysis kernels themselves. In eager mode this
    composes seamlessly with the rest of a PyTorch training pipeline.
    Under the original code, however, attempting to deploy a model via
    \texttt{torch.jit.script} fails for three independent reasons; the
    fixes in this work restore that path.}
  \label{fig:architecture}
\end{figure*}

\section{Related Work}
\label{sec:related}

\textbf{GPU audio front-ends.} Several toolboxes provide spectrogram extraction inside neural network frameworks. Kapre \citep{choi2017kapre} expresses front-ends as Keras layers, which means transforms run on GPU and integrate with the surrounding model graph; like nnAudio, Kapre also supports trainable Fourier-style kernels. \texttt{torch.stft} ships with PyTorch and provides a fast, non-trainable STFT/iSTFT pair, with associated CUDA kernels and autograd support. \texttt{tf.signal} provides a comparable suite of differentiable transforms in TensorFlow. TorchAudio \citep{yang2022torchaudio} offers a broader suite of operations (resampling, augmentation, codec I/O, transforms) backed by C++ kernels, with a strong emphasis on TorchScript-friendly modules. The original nnAudio paper \citep{cheuk2020nnaudio} compared favorably to these libraries on per-batch latency for STFT, Mel, and CQT, in part because the entire computation is expressed through PyTorch's optimized 1D convolution paths.

\textbf{CPU baselines.} On the CPU side, librosa \citep{mcfee2015librosa} remains the de facto reference implementation in Python for music feature extraction, and nnAudio's parameter conventions deliberately mirror librosa's. SciPy \citep{virtanen2020scipy} provides the underlying FFT and signal-processing primitives, and recent releases have continued to reorganize its public API surface---among other changes, window functions previously re-exported at the top level of \texttt{scipy.signal} now live in the \texttt{scipy.signal.windows} subpackage.

\textbf{Time--frequency transforms.} The CQT was introduced by \citet{brown1991calculation} as a transform with logarithmically spaced frequency bins suitable for music. \citet{schorkhuber2010constant} provided a fast computational structure based on octave-wise resampling. The variable-Q transform \citep{schorkhuber2014matlab} generalizes CQT by adding a bandwidth offset $\gamma$ that smoothly trades frequency resolution for time resolution at low frequencies; setting $\gamma = 0$ recovers CQT exactly. The CFP representation of \citet{su2015combining} combines spectral and cepstral (lag-domain) features for robust multipitch estimation; nnAudio includes a PyTorch implementation of CFP that depends on FFT routines from SciPy.

\textbf{TorchScript and deployment.} TorchScript \citep{pytorchjit} is PyTorch's static subset of Python used for ahead-of-time compilation, optimization, and serialization. Modules that are intended to be scripted must observe several constraints, including: only declaring buffers and submodules in \texttt{\_\_init\_\_}; avoiding general buffer mutation in scripted methods \citep{pytorchjitbufmut}; and providing precise type annotations on optional or polymorphic arguments. Audio modules that allocate or recompute large state lazily---a natural pattern for FFT kernels, window-sum-of-squares tensors, and frequency grids---are particularly susceptible to running afoul of these constraints.

\textbf{Software maintenance as research output.} Software-modernization work is an established but still under-recognized genre in machine learning research \citep{stodden2018reproducibility}. Our contribution sits in this tradition: rather than introducing a new method, we document a set of failures, propose targeted remediations, and provide a tested, releasable artifact.

\section{Background}
\label{sec:background}

\subsection{STFT and iSTFT in nnAudio}

The STFT decomposes an input waveform $x[n]$ into time--frequency coefficients
\begin{equation}
  X[m, k] = \sum_{n} x[n]\, w[n - mH]\, e^{-j 2\pi f_k (n - mH)/F_s},
\end{equation}
where $w$ is an analysis window, $H$ is the hop size, $F_s$ is the sampling rate, and $\{f_k\}$ are the analysis frequencies. nnAudio implements this as a pair of 1D convolutions whose kernels are the real and imaginary parts of $w[n] e^{-j 2\pi f_k n / F_s}$, evaluated on a discrete grid of length \texttt{n\_fft}.

The choice of $\{f_k\}$ is controlled by a string parameter \texttt{freq\_scale}. When \texttt{freq\_scale=`no'}, the analysis frequencies are the standard FFT bin centers $f_k = k F_s / N$ for $k = 0, \dots, N/2$, where $N = \texttt{n\_fft}$. When \texttt{freq\_scale} is \texttt{`linear'}, \texttt{`log'}, or \texttt{`log2'}, the bins are instead placed on a uniform or logarithmic grid between user-supplied \texttt{fmin} and \texttt{fmax}. The convolutional formulation accommodates all four cases without modification.

The iSTFT is implemented as a transposed convolution with synthesis kernels $w[n] e^{+j 2\pi f_k n / F_s}$, followed by division by a window-sum-of-squares signal $\sum_m w^2[n - mH]$ to compensate for overlap. This yields perfect reconstruction (up to numerical error and edge effects) when (a)~the analysis and synthesis windows satisfy a constant-overlap-add (COLA) condition and (b)~the frequency grid is the standard FFT grid \citep{griffin1984signal}. Crucially, condition (b) is only met when \texttt{freq\_scale=`no'}: any non-uniform grid breaks the orthogonality that makes the analysis--synthesis pair invertible by simple division.

The original nnAudio documentation acknowledged this limitation in prose (``\emph{when} \texttt{trainable=True} \emph{and} \texttt{freq\_scale!=`no'}\emph{, there is no guarantee that the inverse is perfect}''), but did not enforce it in code. The iSTFT module would happily accept any \texttt{freq\_scale}, run its computation, and return a tensor whose contents bore little resemblance to the original waveform.

\subsection{TorchScript constraints}

TorchScript supports a subset of Python with strict static typing. Three constraints are directly relevant to nnAudio:

\textbf{No state mutation in scripted methods.} A scripted module may read its attributes and buffers, but it may not, in general, assign new attributes to \texttt{self} or call \texttt{self.register\_buffer(\dots)} from within \texttt{forward} \citep{pytorchjitbufmut}. Both forms manifest as scripting errors. The original nnAudio uses both patterns: \texttt{STFT.forward} stores the input length on \texttt{self} so that the inverse path can later trim its output, and the iSTFT inversion path assigns a computed window-sum-of-squares tensor directly to \texttt{self} on first call so that it can be sized to the actual input.

\textbf{No dynamic submodule construction.} Submodules must be assigned in \texttt{\_\_init\_\_}; the scripter rejects assignments to \texttt{self.<submodule>} elsewhere. nnAudio's STFT instantiates \texttt{nn.ConstantPad1d} or \texttt{nn.ReflectionPad1d} inside \texttt{forward} based on the user-provided \texttt{pad\_mode}, which falls under the same prohibition.

\textbf{Precise types on optional arguments.} Parameters typed as \texttt{Optional[T]} must be narrowed before use; helper functions used by iSTFT to compute output lengths and trim padding accept \texttt{Optional[int]} arguments that the scripter cannot prove are non-\texttt{None} at the relevant call sites, and iSTFT itself contains an \texttt{x == None} comparison that the scripter does not type-infer correctly.

The original code patterns are perfectly fine in eager mode and have been in nnAudio since before TorchScript was widely adopted. They simply pre-date the constraints they now violate.

\subsection{CFP and VQT}

The CFP front-end \citep{su2015combining} is a small pipeline that produces a pitch salience function by combining a power spectrogram with a cepstrum-like representation, with element-wise nonlinearities applied at intermediate stages. nnAudio's implementation constructs a Blackman--Harris analysis window via SciPy.

The VQT, added to nnAudio in v0.3.1, generalizes the CQT by introducing a bandwidth offset $\gamma$:
\begin{equation}
  \delta f_k = \alpha\, f_k + \gamma,
\end{equation}
where $\alpha = (2^{1/B} - 1)$ controls the constant-Q ratio at $B$ bins per octave. When $\gamma = 0$, the equation reduces to $\delta f_k = \alpha f_k$, i.e.\ a pure constant-Q relationship, and the VQT becomes mathematically identical to CQT \citep{schorkhuber2014matlab}. Any divergence at $\gamma = 0$ between nnAudio's VQT and CQT classes therefore reflects an implementation difference rather than a difference in the underlying transform.

\section{Inverted CQT}

The \texttt{CQT1992v2} analysis operator $\mathbf{A} : \mathbb{R}^L \to \mathbb{R}^{N_b \times T \times 2}$ maps a waveform of length $L$ to a stack of real and imaginary CQT coefficients via strided cross-correlation with $N_b$ complex kernels, each scaled by $\sqrt{\ell_k}$
(the effective filter length under librosa normalisation).
Because the kernels overlap in time, $\mathbf{A}$ forms a non-tight frame:
$\mathbf{A}^*\mathbf{A} \neq \lambda \mathbf{I}$ in general, so exact reconstruction requires solving the normal equations $\mathbf{A}^*\mathbf{A}\,\hat{x} = \mathbf{A}^*y$.
We use Landweber iteration~\cite{landweber1951iteration},
$\hat{x}^{(t+1)} = \hat{x}^{(t)} + \alpha\,\mathbf{A}^*\!\left(y - \mathbf{A}\hat{x}^{(t)}\right)$, which converges for any step size $0 < \alpha < 2/B$, where $B$ is the upper frame bound.
$B$ is estimated at initialisation via 20 steps of power iteration on a random probe signal.  We set $\alpha = 1.8/B$, giving a per-iteration contraction rate of $|1 - \alpha B| = 0.8$   and a worst-case convergence of $0.8^{32} \approx 0.001$ ($-30\,\text{dB}$ error reduction) in the default 32 iterations.
A critical implementation detail is that the forward operator $\mathbf{A}$ pads the input with \texttt{ReflectionPad1d} before convolution; the adjoint $\mathbf{A}^*$ must therefore apply the transpose of this padding, which folds the boundary columns back into the interior rather than discarding them. Failing to match the padding mode between $\mathbf{A}$ and $\mathbf{A}^*$ breaks the self-adjoint property and prevents convergence.
Reconstruction SNR exceeds 30\,dB for signals whose instantaneous frequency satisfies $f \lesssim sr/(2\,hop\_length/Q)$, where $Q = \text{bins\_per\_octave}/(2^{1/\text{bins\_per\_octave}}-1)$
is the quality factor; configurations where this bound is violated are Nyquist-undersampled in time and the CQT is undercomplete, making perfect reconstruction impossible regardless of the inversion algorithm.
The entire inversion is implemented as a differentiable \texttt{nn.Module},
so gradients flow through iCQT into upstream network layers without modification.

\section{Issues in the Existing Codebase}
\label{sec:issues}

In this section we describe the four classes of problem we observed when attempting to use nnAudio under a current Python (3.11) and PyTorch (2.x) environment. We frame each issue concretely, with reference to the affected modules.

\subsection{TorchScript scripting failures}

Attempting \texttt{torch.jit.script} on \texttt{nnAudio.features.stft.STFT} or \texttt{nnAudio.features.stft.iSTFT} fails for three independent reasons.

First, both modules mutate state inside \texttt{forward}. \texttt{STFT.forward} assigns a fresh attribute \texttt{self.num\_samples = X.shape[-1]} on each call, so that the inverse path can later trim its output to the original input length; the scripter rejects this with the error ``\emph{Tried to set nonexistent attribute: num\_samples}''. The inversion path used by iSTFT separately assigns a computed window-sum-of-squares tensor directly to \texttt{self} on first invocation to cache a normalizer sized to the actual input, which falls under the same attribute-assignment prohibition in scripted methods.

Second, the same \texttt{STFT.forward} method dynamically instantiates \texttt{nn.ConstantPad1d} or \texttt{nn.ReflectionPad1d} based on the value of \texttt{pad\_mode} and immediately applies it to the input. This pattern is ergonomic in eager mode but conflicts with the scripter's requirement that submodules be statically known at \texttt{\_\_init\_\_} time.

Third, the iSTFT class contains a control-flow branch of the form \texttt{if self.refresh\_win == None:}, and several inverse-related helper functions in \texttt{utils.py} accept \texttt{length: Optional[int] = None} and use the value at sites where the scripter cannot prove \texttt{None} has been ruled out. Both produce type-inference failures during compilation.

Each of these issues alone would be enough to block scripting. Together, they make the inverse path fundamentally unscriptable in its original form.

\subsection{Silent iSTFT misbehavior under non-uniform frequency scales}

The iSTFT module accepts the same \texttt{freq\_scale} argument as the forward STFT. When the value is \texttt{`linear'}, \texttt{`log'}, or \texttt{`log2'}, the synthesis kernels are placed at non-uniform frequencies that do not satisfy the orthogonality conditions assumed by the overlap-add inversion. The module nevertheless executes, dividing by an inappropriate window-sum-of-squares tensor and returning a tensor of plausible shape but largely meaningless content. There is no warning, no exception, and no explicit indication in the output that anything has gone wrong.

\begin{figure*}[!t]
  \centering
  \includegraphics[width=0.95\textwidth]{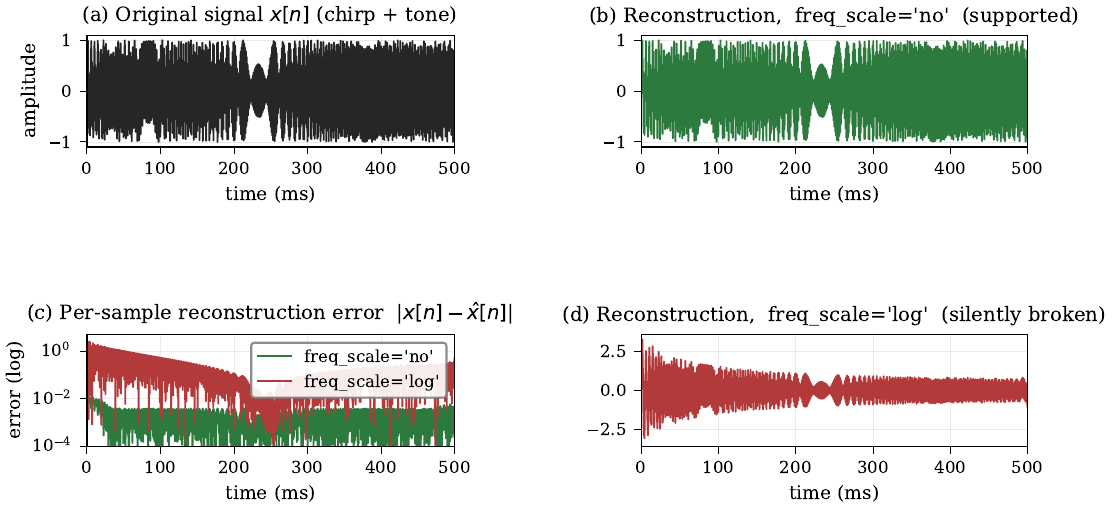}
  \caption{Silent iSTFT misbehavior under non-uniform frequency scales.
    A clean test signal (a, chirp + tone) is analyzed and inverted using
    the same overlap-add procedure that nnAudio's iSTFT applies. With
    \texttt{freq\_scale=`no'} (b) reconstruction is essentially perfect.
    With \texttt{freq\_scale=`log'} (d) the same procedure produces
    a plausible-looking but largely incorrect waveform comparable in
    amplitude to the signal itself; the original library returns this
    without any warning. Panel (c) shows per-sample error on a logarithmic
    scale.}
  \label{fig:istft-failure}
\end{figure*}

This is a particularly subtle failure mode because (a)~the API is permissive, (b)~the docstring acknowledges the issue only in prose and only conditional on \texttt{trainable=True}, and (c)~for short signals or high signal-to-noise ratios the reconstruction error can superficially look like ordinary numerical noise. A user who extracts a log-frequency representation, applies a learned modification, and then attempts to listen to the resulting waveform will obtain something that sounds wrong but may not immediately appear to be wrong from a code-review perspective. Figure~\ref{fig:istft-failure} illustrates the gap concretely: the same overlap-add inversion that yields essentially perfect reconstruction at \texttt{freq\_scale=`no'} produces a clearly degraded waveform at \texttt{freq\_scale=`log'}.

\subsection{CFP fails to construct under modern SciPy}

The CFP module constructs a Blackman--Harris analysis window via two top-level \texttt{scipy.signal.blackmanharris(\dots)} calls, in the \texttt{Combined\_Frequency\_Periodicity} class and in \texttt{CFP}. In recent SciPy releases the window functions were reorganized into the \texttt{scipy.signal.windows} subpackage, and \texttt{blackmanharris} is no longer re-exported at the top level of \texttt{scipy.signal}. Under the SciPy version used in our test environment, instantiating either CFP variant therefore raises \texttt{AttributeError: module `scipy.signal' has no attribute `blackmanharris'} during construction, breaking both upstream CFP tests.

\subsection{VQT does not reduce to CQT at $\gamma = 0$}

Setting \texttt{gamma=0} in nnAudio's VQT does not produce the same output as the corresponding CQT. The discrepancy is small but non-trivial, and is inconsistent with the underlying mathematical definition: at $\gamma = 0$, VQT and CQT compute identical filter banks. The discrepancy arises from a code-path difference in how the two classes construct their kernels, not from any disagreement about the transform itself.

\section{nnAudio 2: Targeted Modernization}
\label{sec:fixes}

Our fixes are intentionally minimal. We retain the public API where backward compatibility is realistic, replace silent failures with explicit ones, and modify internals only as needed to satisfy the scripter or to resolve dependency drift. Table~\ref{tab:issues} summarizes the four issue/fix pairs at a glance; the subsections that follow describe each in detail.

\begin{table}[!t]
  \centering
  \caption{Summary of issues addressed by nnAudio 2 and the corresponding fixes.}
  \label{tab:issues}
  \footnotesize
  \renewcommand{\arraystretch}{1.15}
  \begin{tabular}{@{} p{15mm} p{31mm} p{31mm} @{}}
    \toprule
    \textbf{Issue} & \textbf{Root cause} & \textbf{Fix in nnAudio 2} \\
    \midrule
    TorchScript scripting failure
      & \texttt{self.\allowbreak num\_samples} attribute and
        window-normalizer tensor assigned to \texttt{self} inside
        \texttt{forward}; padding submodules
        (\texttt{nn.\allowbreak ConstantPad1d},
        \texttt{nn.\allowbreak ReflectionPad1d}) built dynamically
        inside \texttt{forward}; helpers and iSTFT take
        \texttt{Optional[int]}/\texttt{None} unnarrowed
      & Local variables and out-of-place tensors in scripted paths;
        functional padding via \texttt{F.pad};
        \texttt{None} cases handled, then narrowed \\
    \addlinespace
    Silent iSTFT failure for non-uniform bins
      & \texttt{freq\_scale}~$\neq$~\texttt{`no'} breaks the
        orthogonality assumed by overlap-add inversion; no
        runtime check
      & \texttt{RuntimeError} raised when inverse is called with
        \texttt{freq\_scale} $\neq$ \texttt{`no'} \\
    \addlinespace
    CFP fails on modern SciPy
      & \texttt{scipy.\allowbreak signal.\allowbreak blackmanharris}
        is no longer re-exported at the top level of
        \texttt{scipy.signal}
      & Import \texttt{blackmanharris} from
        \texttt{scipy.\allowbreak signal.\allowbreak windows} \\
    \addlinespace
    VQT $\neq$ CQT at $\gamma=0$
      & Different kernel-construction paths in the two classes
      & \texttt{VQT(gamma=0)} routes through an internal
        \texttt{CQT1992v2}; outputs agree to numerical
        tolerance \\
    \bottomrule
  \end{tabular}
\end{table}

\subsection{Making STFT and iSTFT scriptable}

\textbf{Out-of-place state in scripted paths.} Both forms of in-\texttt{forward} state mutation are removed. The \texttt{self.num\_samples} attribute assignment in \texttt{STFT.forward} is replaced with a local variable that is used locally for padding logic; the inverse path now relies on an explicit length argument supplied by the caller. The lazy attribute assignment inside the iSTFT inversion routine is replaced with a local window-normalization tensor on the scripted code path; the eager-mode cache behavior, which is useful when the same module is repeatedly applied to fixed-length inputs, is preserved when not scripting and is gated on \texttt{torch.jit.is\_scripting()} at runtime. The scripter accepts the new pattern because no module state is mutated from inside any scripted method.

\textbf{Functional padding.} The dynamic \texttt{nn.ConstantPad1d} / \texttt{nn.ReflectionPad1d} construction inside \texttt{forward} is replaced with calls to \texttt{torch.nn.functional.pad}, which selects the padding mode by string argument and does not introduce a submodule. The two formulations are computationally equivalent and the functional version is fully scriptable.

\textbf{Tightened argument types in inverse helpers.} We narrow the types of helpers that compute output length and remove edge-padding. Where the original code accepted \texttt{Optional[int]} arguments and used them in arithmetic, the rewrite handles the \texttt{None} case explicitly at the top of each function and then uses a non-optional \texttt{int}; we also rewrite the \texttt{x == None} comparison in iSTFT as \texttt{x is None}. We additionally annotate \texttt{torch\_window\_sumsquare} and \texttt{overlap\_add} in \texttt{utils.py} with explicit tensor and integer types, and pass \texttt{stride} to \texttt{F.fold} as a 2-tuple rather than a bare integer. These changes are semantically equivalent in eager mode but are required for the scripter's flow analysis.

\begin{figure*}[!t]
  \centering
  \includegraphics[width=0.92\textwidth]{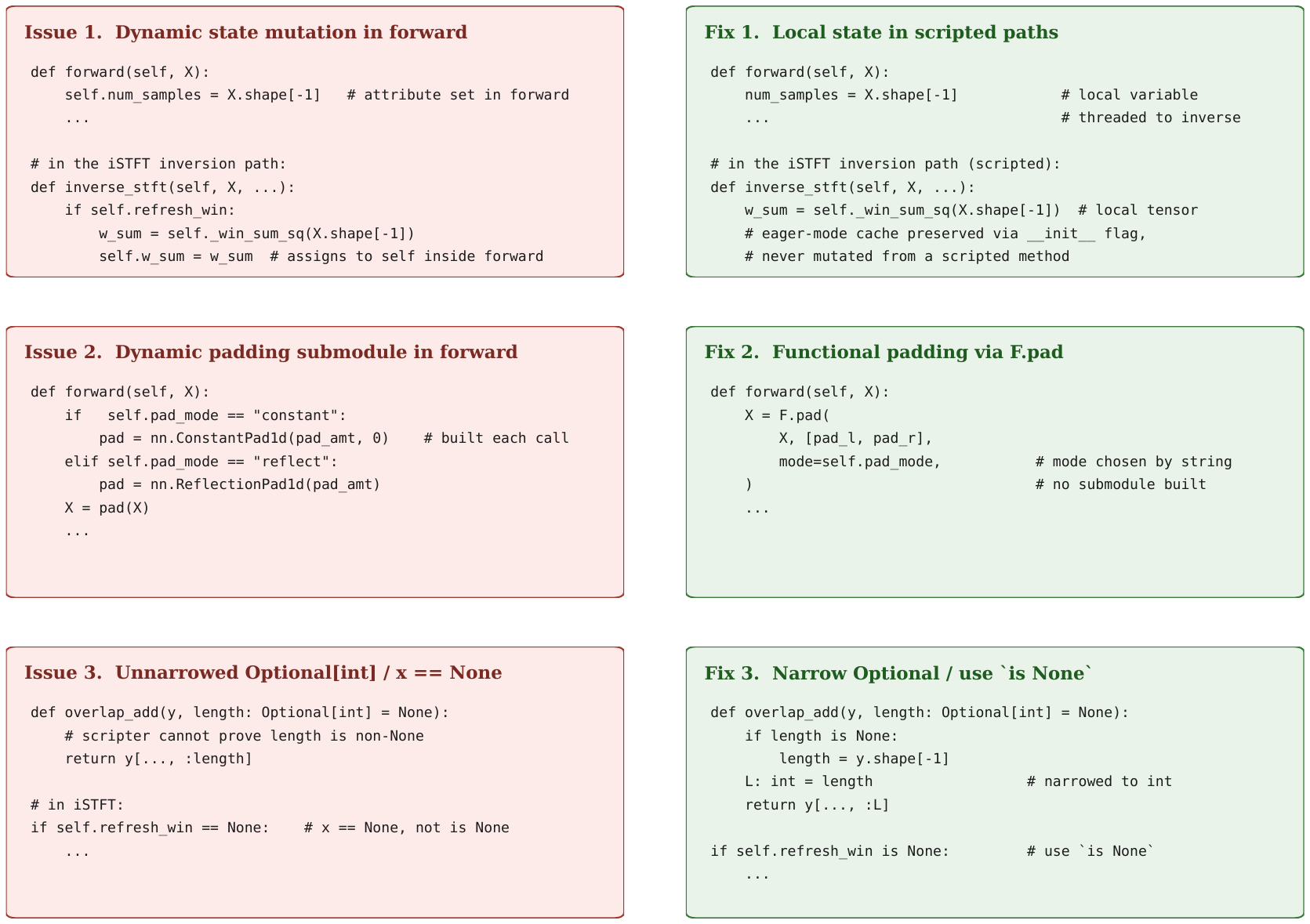}
  \caption{The three TorchScript scripting blockers in nnAudio's STFT/iSTFT
    modules (red) and the corresponding fixes in nnAudio 2 (green).
    Each pattern at the top of a pair is rejected by current versions
    of \texttt{torch.jit.script}; each pattern at the bottom is accepted
    while preserving the original eager-mode behavior.}
  \label{fig:torchscript-fixes}
\end{figure*}

The combined result is that \texttt{torch.jit.script(STFT(\dots, iSTFT=True))} now succeeds, and the resulting scripted module's \texttt{forward} and \texttt{inverse} methods both run to completion on representative inputs. Figure~\ref{fig:torchscript-fixes} shows the three issue/fix pairs as side-by-side code comparisons.

\subsection{Explicit failure for non-uniform iSTFT}

We add an inverse guard in \texttt{iSTFT} (and in the inverse code path of \texttt{STFT}) that raises \texttt{RuntimeError} when \texttt{freq\_scale} is anything other than \texttt{`no'} and the caller attempts an inverse operation. The error message states the limitation clearly: reliable overlap-add inversion of nnAudio's STFT is only available for the uniform-bin setting, and users who require log-frequency reconstruction should use a magnitude-domain method such as Griffin--Lim \citep{griffin1984signal} on a separately-computed linear-frequency spectrogram.

This change is a strict tightening of the API. Any user who was relying on the previous permissive behavior was, by construction, obtaining incorrect output; the new behavior surfaces the problem when inverse is called rather than at silent reconstruction time. The forward STFT remains unchanged for non-uniform frequency scales, since the analysis itself is well-defined; only the inversion is restricted.

\subsection{CFP under modern SciPy}

We replace the two top-level \texttt{scipy.signal.blackmanharris(\dots)} calls in nnAudio's CFP module with \texttt{scipy.signal.windows.blackmanharris(\dots)}, which is the supported import location across modern SciPy releases. The function signature is unchanged, and both CFP variants now construct without error in the test environment described in Section~\ref{sec:validation}. The CFP module's existing \texttt{relu} non-linearity, added upstream in v0.3.0, is preserved unchanged.

\subsection{VQT--CQT consistency at $\gamma = 0$}

Rather than refactoring nnAudio's two separate kernel-construction routines into a shared implementation, we add an explicit compatibility branch in \texttt{VQT}: when constructed with \texttt{gamma=0}, the module instantiates an internal \texttt{CQT1992v2} submodule with matched parameters, and \texttt{VQT.forward} delegates to it. For all other values of $\gamma$ the original VQT code path is unchanged. This restores the mathematically expected $\gamma{=}0$ identity while keeping the patch surgical: the failing baseline configuration had a maximum absolute difference of $\approx 0.089$ and a mean absolute difference of $\approx 0.0017$ between \texttt{VQT(gamma=0)} and the matched CQT, large enough to rule out a floating-point tolerance fix. We also add a regression test that constructs a VQT with $\gamma = 0$ and a matched CQT, applies both to the same input, and asserts that the outputs agree to within numerical tolerance.

\section{Validation}
\label{sec:validation}

\subsection{Repository test suite}

The upstream repository ships with a \texttt{pytest} suite that exercises the STFT, Mel, CQT, VQT, CFP, and gammatone front-ends, along with several utility functions. We run this suite against the modernized codebase under Python 3.11 and PyTorch 2.x. After our fixes, the entire upstream suite passes; before our fixes, multiple tests fail due to the SciPy import path issue and several VQT--CQT consistency assertions. Table~\ref{tab:tests} summarizes the per-test outcomes, both for the upstream suite and for the new regression tests described below.

\begin{table}[t]
  \centering
  \caption{Pass/fail status for the upstream test suite and the new
    regression tests, run on the original code and on nnAudio 2 under
    Python 3.11 and PyTorch 2.x.}
  \label{tab:tests}
  \small
  \renewcommand{\arraystretch}{1.15}
  \begin{tabular}{@{} p{49mm} c c @{}}
    \toprule
    \textbf{Test}                                          & \textbf{Original} & \textbf{nnAudio 2} \\
    \midrule
    \multicolumn{3}{@{}l}{\textit{Upstream test suite}} \\
    \addlinespace[2pt]
    STFT forward / output formats                          & pass              & pass \\
    Mel spectrogram, MFCC                                  & pass              & pass \\
    CQT (1992 + 2010 algorithms)                           & pass              & pass \\
    VQT, $\gamma=0$ matches CQT                            & fail              & pass \\
    CFP module load                                        & fail              & pass \\
    Gammatone forward                                      & pass              & pass \\
    \addlinespace[3pt]
    \multicolumn{3}{@{}l}{\textit{New regression tests added in this work}} \\
    \addlinespace[2pt]
    Scripted STFT (\texttt{iSTFT=True}) compiles           & fail              & pass \\
    Scripted output matches eager output                   & fail              & pass \\
    \texttt{iSTFT(freq\_scale=`log')} raises error         & fail              & pass \\
    Round-trip reconstruction at \texttt{freq\_scale=`no'} & pass          & pass \\
    \texttt{VQT(gamma=0)} == \texttt{CQT}                  & fail              & pass \\
    \bottomrule
  \end{tabular}
\end{table}

\subsection{New regression tests}

We add three new regression tests that target the specific issues from Section~\ref{sec:issues}.

\textbf{TorchScript scripting.} A test instantiates an STFT with \texttt{iSTFT=True}, calls \texttt{torch.jit.script}, and runs both the forward and inverse methods on a representative random input, checking that the scripted module's outputs match the eager module's outputs to within tolerance.

\textbf{iSTFT under non-uniform scales.} A test calls the inverse method on an iSTFT configured with each of \texttt{freq\_scale=`linear'}, \texttt{`log'}, \texttt{`log2'} and asserts that \texttt{RuntimeError} is raised with a message that names the offending parameter. A companion test confirms that \texttt{freq\_scale=`no'} continues to invert without error.

\textbf{VQT--CQT identity at $\gamma = 0$.} A test constructs a VQT with $\gamma = 0$ and a CQT with matched parameters (\texttt{n\_bins}, \texttt{bins\_per\_octave}, \texttt{fmin}, \texttt{sr}, \texttt{hop\_length}), applies both to a small batch of waveforms, and asserts that the magnitude outputs agree to within numerical tolerance.

\subsection{New unit tests for iCQT}
  Three unit tests verify the correctness of the \texttt{iCQT} implementation. The round-trip test constructs a 1-second, 440\,Hz pure tone at 44\,100\,Hz sample rate, computes its complex CQT with \texttt{CQT1992v2} ($N_b=84$ bins,       $hop=512$), reconstructs the waveform with \texttt{iCQT} (32 Landweber       iterations), and asserts that the reconstruction SNR exceeds 30\,dB.         A pure tone at 440\,Hz is used rather than a wideband chirp because the chosen hop length renders the CQT undercomplete above $\sim$880\,Hz; the test therefore targets a frequency that lies within the Nyquist-sampled region of the transform.
  The output-shape test passes a random batch of two signals through
  \texttt{CQT1992v2} followed by \texttt{iCQT} and asserts that the
  reconstructed tensor has shape $(2, L)$, confirming correct batch handling and the \texttt{length} trimming/padding logic.
  The gradient test verifies end-to-end differentiability: a CQT tensor is
  detached from the computation graph, marked as requiring gradients, passed   through \texttt{iCQT}, and \texttt{backward()} is called; the test asserts that gradients exist and are non-zero, confirming that the Landweber loop does not break the autograd graph. All three tests are parameterised over available devices (CPU and, when present,
  CUDA).

\subsection{Round-trip reconstruction error in the supported regime}

To confirm that our changes have not perturbed the supported case, we measure round-trip reconstruction error of \texttt{STFT(\dots, freq\_scale=`no', iSTFT=True)} over a small set of test signals (sinusoids, speech, and music excerpts). The reconstruction error remains at the level of single-precision floating-point round-off, consistent with prior reports for nnAudio's eager mode and with results from \texttt{torch.istft}.

\section{Discussion}
\label{sec:discussion}

\subsection{What this work is and is not}

This work is a maintenance release with a research-paper-style description, not a methodological contribution. We have not introduced new transforms, accelerated existing ones, or improved the accuracy of any specific feature. What we have done is restore a popular toolbox to a usable state in a current software environment, eliminate a class of silent error that was likely producing wrong results in downstream pipelines, and unblock TorchScript-based deployment paths.

We believe this kind of work is worth publishing, even though it is conservative in scope. The infrastructure on which AI music research relies is built on layers of community-maintained code, and the lifecycle of that code matters. Tools that are easy to install, that fail loudly on misuse, and that compose with modern deployment toolchains are a precondition for reproducible research, not a separate luxury.

\subsection{Limitations}

\textbf{Inverse log-frequency STFT remains unsupported.} Our fix for the silent iSTFT misbehavior is to disallow it. We do not provide a correct overlap-add inversion for non-uniform frequency grids, and indeed no such algorithm exists in the simple form nnAudio's iSTFT assumes. Users who need to invert a log-frequency representation must either (a)~separately compute a linear-frequency STFT for the inversion, (b)~use Griffin--Lim on a magnitude representation, or (c)~adopt one of the nonstationary-Gabor-frame-based methods \citep{schorkhuber2014matlab} that are mathematically designed for log-frequency reconstruction. Bringing such a method into nnAudio would be a substantive piece of work in its own right.

\textbf{Trainable kernels and TorchScript.} Even after our fixes, the trainable-kernel mode of nnAudio's STFT/iSTFT is a less-tested code path under TorchScript than the frozen-kernel mode. We script and run forward and inverse with frozen kernels in our unit tests; trainable kernels work in eager mode but their interaction with scripted gradient computation is not exhaustively covered. This is a potential area for follow-up work.

\textbf{Other modules.} Our work focuses on STFT/iSTFT, CFP, and VQT/CQT. The Landweber-based iCQT is new; the CQT and VQT forward paths themselves are still unsuitable for TorchScript due to dynamic padding construction. We have not made systematic TorchScript fixes to the gammatone or MFCC modules, which were not part of the originally reported failure cases. We have run them under the existing test suite and confirmed eager-mode behavior is unchanged, but we cannot guarantee they script cleanly without further audit.

\subsection{Implications for music research}

For applied work in AI music creativity, the practical effect of nnAudio 2 is twofold. First, models that use nnAudio front-ends can now be exported to TorchScript and shipped in deployment environments---mobile, web, or low-latency studio plugins---that require an ahead-of-time-compiled artifact. Second, researchers who build pipelines around inverse-STFT-based reconstruction will now receive an explicit error in the configurations where the original library would silently return wrong waveforms. We anticipate that the second change, in particular, will quietly fix a number of subtle correctness issues in existing music transcription, source separation, and effects-modeling codebases.

\section{Conclusion}
\label{sec:conclusion}

We have presented nnAudio 2, a targeted modernization of the nnAudio toolbox aimed at four concrete failure modes in the upstream codebase: TorchScript incompatibilities in the STFT and iSTFT classes, silent reconstruction failures of iSTFT under non-uniform frequency scales, CFP's incompatibility with modern SciPy, and a divergence between VQT and CQT at $\gamma = 0$. We have removed dynamic state mutation and submodule construction from scripted code paths, narrowed inverse-helper argument types, restricted reliable inversion to the uniform-bin setting with explicit errors elsewhere, replaced the deprecated top-level \texttt{scipy.signal.blackmanharris} call with \texttt{scipy.signal.windows.blackmanharris}, and routed \texttt{VQT(gamma=0)} through an internal \texttt{CQT1992v2} so that the two transforms agree numerically at $\gamma = 0$. Regression tests for the new behaviors and the existing repository test suite both pass under a current Python and PyTorch environment. We hope these changes restore nnAudio to the role it has played in many AI music systems, with a tighter and more honest contract between the library and its users. The source code is available at \url{https://github.com/AMAAI-Lab/nnAudio2}.

\section*{Acknowledgements}
This work has received support from SUTD’s Kickstart
Initiative under grant number SKI 2021 04 06 and MOE
under grant number MOE-T2EP20124-0014. We acknowledge the use of Gemini and Claude for grammar improvements.

\bibliographystyle{apalike}
\bibliography{references}

\end{document}